\documentclass[conference]{IEEEtran}
\IEEEoverridecommandlockouts
\usepackage{amsmath,amssymb}

\usepackage{comment}
\usepackage{svg}
\setsvg{svgpath=02_figures/}
\usepackage{multirow}
\usepackage{booktabs} 

\usepackage{etoolbox}
\makeatletter
\patchcmd{\@makecaption}
  {\scshape}
  {}
  {}
  {}
\makeatother

\def\BibTeX{{\rm B\kern-.05em{\sc i\kern-.025em b}\kern-.08em
    T\kern-.1667em\lower.7ex\hbox{E}\kern-.125emX}}
\usepackage{mathtools}
\DeclarePairedDelimiter\ceil{\lceil}{\rceil}
\usepackage{acronym}
\acrodef{eeg}[EEG]{electroencephalogram}
\acrodef{cnn}[CNN]{Convolutional neural network}
\acrodef{bmi}[BMI]{brain--machine interface}
\acrodef{bci}[BCI]{Brain--computer interface}
\acrodef{mi}[MI]{motor-imagery}
\acrodef{mm}[MM]{motor-movement}
\acrodef{smr}[SMR]{sensory motor rhythm}
\acrodef{csp}[CSP]{common spatial pattern}
\acrodef{mcu}[MCU]{microcontroller unit}
\acrodef{sstl}[SS-TL]{subject-specific transfer learning}
\acrodef{cv}[CV]{cross-validation}
\acrodef{lr}[LR]{learning rate}
\acrodef{dr}[DR]{dropout rate}
\acrodef{hp}[HP]{hyperparameter}
\acrodef{sgd}[SGD]{stochastic gradient descent}
\acrodef{dsp}[DSP]{digital signal processor}
\acrodef{fpu}[FPU]{floating-point unit}
\acrodef{fbcsp}[FBCSP]{filter bank common spatial pattern}
\acrodef{svm}[SVM]{support vector machine}
\acrodef{lda}[LDA]{linear discriminant analysis}
\acrodef{mops}[MOPS]{million operations per second}
\acrodef{soa}[SoA]{state-of-the-art}
\acrodef{elu}[ELU]{exponential linear unit}
\acrodef{macc}[MACC]{multiply-and-accumulate}

\renewcommand{\baselinestretch}{0.965}

\usepackage{fancyhdr}
\fancypagestyle{mahmood}{%
  \fancyhf{} 
  
  \fancyhead[C]{\footnotesize \textcopyright 2020 IEEE. Personal use of this material is permitted.  Permission from IEEE must be obtained for all other uses, in any current or future media, including reprinting/republishing this material for advertising or promotional purposes, creating new collective works, for resale or redistribution to servers or lists, or reuse of any copyrighted component of this work in other works.}
}%
\makeatletter
\let\ps@IEEEtitlepagestyle\ps@mahmood
\makeatother

\begin{document}

\title{An Accurate EEGNet-based Motor-Imagery Brain--Computer Interface for Low-Power Edge Computing\vspace{-0.4cm}}

\author{\IEEEauthorblockN{
    Xiaying Wang\IEEEauthorrefmark{1}\IEEEauthorrefmark{2}, 
    Michael Hersche\IEEEauthorrefmark{1}\IEEEauthorrefmark{2}, 
    Batuhan T{\"o}mekce\IEEEauthorrefmark{2}, 
    Burak Kaya\IEEEauthorrefmark{2}, 
    Michele Magno\IEEEauthorrefmark{2}\IEEEauthorrefmark{3},
    Luca Benini\IEEEauthorrefmark{2}\IEEEauthorrefmark{3}}
    \IEEEauthorblockA{\\[-2mm]\IEEEauthorrefmark{2}ETH Z{\"u}rich, Dept. EE \& IT,  Switzerland \hspace{15mm}
    \IEEEauthorrefmark{3}University of Bologna, DEI, Italy}
    \vspace{-0.7cm}
    \thanks{\IEEEauthorrefmark{1}X. Wang and M. Hersche contributed equally to this work as first authors. Corresponding emails: \{xiaywang, herschmi\}@iis.ee.ethz.ch}
    }

\maketitle

\begin{abstract}
%
This paper presents an accurate and robust embedded motor-imagery brain--computer interface (MI-BCI). The proposed novel model, based on EEGNet~\cite{Lawhern2018EEGNet:Interfaces}, matches the requirements of memory footprint and computational resources of low-power microcontroller units (MCUs), such as the ARM Cortex-M family. 
Furthermore, the paper presents a set of methods, including temporal downsampling, channel selection, and narrowing of the classification window, to further scale down the model to relax memory requirements with negligible accuracy degradation. 
Experimental results on the Physionet EEG Motor Movement/Imagery Dataset show that standard EEGNet achieves 82.43\%, 75.07\%, and 65.07\% classification accuracy on 2-, 3-, and 4-class MI tasks in global validation, outperforming the state-of-the-art (SoA) convolutional neural network (CNN) by 2.05\%, 5.25\%, and 6.49\%.
Our novel method further scales down the standard EEGNet at a negligible accuracy loss of 0.31\% with 7.6$\times$ memory footprint reduction and a small accuracy loss of 2.51\% with 15$\times$ reduction.
The scaled models are deployed on a commercial Cortex-M4F MCU taking 101\,ms and consuming 4.28\,mJ per inference for operating the smallest model, and on a Cortex-M7 with 44\,ms and 18.1\,mJ per inference for the medium-sized model, enabling a fully autonomous, wearable, and accurate low-power BCI.
\end{abstract}

\begin{IEEEkeywords}
Brain--computer interface, motor-imagery, CNN, embedded systems, edge computing
\end{IEEEkeywords}

\section{Introduction}
\acp{bci} aim to provide a communication and control channel based on the recognition of the subject’s intentions, e.g., when performing \ac{mi}, from neural activity typically recorded by noninvasive \ac{eeg}
electrodes~\cite{Graimann2009BrainComputerIntroduction}. 
\ac{mi}-\ac{bci} systems are designed to find patterns in the \ac{eeg} signals and match the signal to the motor motion that was imagined by the subject.
Such information could enable communication for severely paralyzed users, control of a wheelchair~\cite{Yu2017Self-PacedPotential}, or assistance in stroke rehabilitation~\cite{Frolov2017Post-strokeTrial.}. 

\ac{mi}-\acp{bci} are still susceptible to errors mostly due to high inter- and intra-subject variance in \ac{eeg} data~\cite{Tangermann2012ReviewIV.,Lotte2018AUpdate}, resulting in low classification accuracy. 
Traditional methods approach this challenge with robust feature extractors, typically \ac{fbcsp}~\cite{Ang2008FilterInterface} or Riemannian covariances~\cite{Yger2017RiemannianReview}, and classify these features with \ac{lda} or \acp{svm}~\cite{Lotte2018AUpdate}.
\acp{cnn} have been proposed as a competitive solution in \ac{eeg} classification, while requiring fewer parameters to learn and being computationally cheaper in inference than traditional \ac{bci} methods~\cite{Schirrmeister2017DeepVisualization,Lawhern2018EEGNet:Interfaces}.
However, today's \ac{cnn} models are designed to be executed on a CPU or GPU, requiring \ac{eeg} data to be transmitted from the sensor node to an external compute engine through wired or wireless communication. 
Due to their computational complexity and resource requirements, those models have been predominantly confined to cloud computing with high-performance computers rather than used in real-world \ac{bci} applications, where latency, privacy, and wearability are crucial requirements besides the accuracy~\cite{Chen2019DeepReview, angrisani2018TIM}.
%
Recently, a new generation of wearable \acp{bci} is attracting the academic and industrial researchers. 
An increasing number of battery-operated wearable solutions, using \acp{mcu}, are proposed to bring computing capabilities towards the ``edge'' to perform real-time near-sensor computation~\cite{Chen2019DeepReview,Guermandi2018APotentials_short,Kartsch2019BioWolf:Connectivity_short, Wang_fann-on-mcu}.
Edge computing and near-sensor computation offer the following advantages: 1) lower energy consumption for the data transmission between sensors and remote processing; 2) longer battery lifetime; 3) significantly shorter latency compared to remote computation; 4) user comfort; 5) security and privacy improvements, as the data are processed locally and only little information is transmitted wirelessly if necessary.

On the other hand, edge computing poses several challenges when it needs to match the requirement of long-term battery operation, mandatory in wearable devices, and to continuously perform complex \ac{bci} models (e.g., \acp{cnn}) with a low-power processor. For instance, the ARM Cortex-M series is the most popular family of low-power processors used in embedded wearable devices~\cite{lai2018cmsis}. Those \acp{mcu} allow several hours lifetime with a small-scale battery, but they have a resource-constrained architecture. For example, an ARM Cortex-M4F processor offers few KB of RAM and \ac{mops} in a power envelope of few mW~\cite{eggimann2019risc}. The more recent ARM Cortex-M7 provides an even better performance up to 300-400\,MOPS with a higher power consumption of few hundred mW. To achieve the goal of deploying complex and accurate \ac{cnn} models on these tiny microprocessors, the models need to be re-thought and redesigned with the above-mentioned constraints in mind. Moreover, many researchers have demonstrated for computer vision that reducing the model size with clever network optimization techniques does not always cause a performance degradation~\cite{lai2018cmsis,iandola2016squeezenet}. 

This paper proposes a novel embedded model for \ac{mi}-\ac{bci} that focuses on bringing the next generation of edge \ac{bci} on autonomous wearable systems. 
The main contributions of the paper are as follows:
\begin{itemize}
    \item We propose a novel embedded \ac{mi}-\ac{bci} model which outperforms the \ac{soa} model on the Physionet EEG Motor Movement/Imagery Dataset~\cite{goldberger2000physiobank}. The model is based on EEGNet architecture~\cite{Lawhern2018EEGNet:Interfaces} and achieves a global validation accuracy of 82.43\%, 75.07\%, and 65.07\% on \mbox{2-,} \mbox{3-,} and 4-class \ac{mi} task, which is 2.05\%, 5.25\%, and 6.49\% higher than the \ac{soa} \ac{cnn}~\cite{Dose2018AnBCIs}. 
    %
    \item 
    We further propose methods to reduce the memory footprint for the execution of the model by temporal downsampling, channel reduction, and narrowing down the time window considered for performing one classification, without significant loss in accuracy. This allows us to target low-power embedded devices with very tight constraints.
    %
    \item We evaluate experimentally the benefits of our model in terms of energy consumption, latency, and accuracy on two different platforms: ARM Cortex-M4F and Cortex-M7. We compare the two platforms executing the inference of 4-class \ac{mi} with accurate measurements. 
    
     
\end{itemize}

To the best of our knowledge, no previous work has evaluated \ac{mi}-\ac{bci} on these low-power \acp{mcu} using \acp{cnn} by considering both runtime and power measurements besides the classification accuracy. Finally, we release open-source code developed in this work\footnote{https://github.com/MHersche/eegnet-based-embedded-bci}.

\section{Related work}\label{sec:related}

The recent literature on \ac{mi}-\acp{bci} is very rich, mostly considering feature extraction and classifiers separately.
EEG signals are typically pre-processed using spectral and spatial filters followed by log-energy feature calculation, better known as \ac{fbcsp}~\cite{Ang2008FilterInterface, Hersche2018_short}. 
The multi-spectral features are classified using either \ac{lda}, regularized \ac{lda}, or \acp{svm}~\cite{Lotte2018AUpdate}.
%

Alternatively, the feature extractor and classifier can be combined and trained simultaneously with a \ac{cnn}. Today, \acp{cnn} are among the most accurate \ac{bci} architectures and demonstrated impressive performance~\cite{Dose2018AnBCIs,Schirrmeister2017DeepVisualization,Lawhern2018EEGNet:Interfaces}. 
%
%
Schirrmeister et al.~\cite{Schirrmeister2017DeepVisualization} provide an elaborate study on \ac{cnn} architectures for \ac{mi}-\ac{bci}, where the small Shallow ConvNet achieves an accuracy of 73.59\% on the 4-class \ac{mi}-\ac{bci} competition IV-2a~\cite{Tangermann2012ReviewIV.}. 
With its temporal and spatial filters followed by square-log activation, Shallow ConvNet can be interpreted as a tunable variant of \ac{fbcsp}. 
%

Due to limited amount of data provided in the \ac{mi}-\ac{bci} competition IV-2a dataset containing recordings of only 9 subjects with 144 \ac{mi}-trials per class, a variant of Shallow ConvNet has been trained and validated in~\cite{Dose2018AnBCIs} on the much larger Physionet EEG Motor Movement/Imagery Dataset~\cite{goldberger2000physiobank}, with recordings of 109 subjects with 21 \ac{mi}-trials per class which is overall $\approx$2$\times$ the amount of \ac{mi}-trials. 
The model consists of around 305 thousand trainable parameters and is trained and validated globally in 5-fold \ac{cv} across subjects.
%
It has achieved \ac{soa} accuracy of 80.38\%, 69.82\%, and 58.58\% on \mbox{2-,} \mbox{3-,} and 4-class \ac{mi} on that dataset. 
Additionally, the global models are adjusted for every subject using \ac{sstl}, which further improved the accuracy by 6.11\%, 9.42\%, and 9.93\%. 
The main differences to the original Shallow ConvNet are the use of ReLU instead of square-log activation and the splitting of the final classification layer into two fully connected layers, which increases the number of trainable parameters.
%
%

Another smaller, yet robust, \ac{cnn} architecture is EEGNet~\cite{Lawhern2018EEGNet:Interfaces}, which achieved the same accuracy as the winner of the BCI competition IV-2a on 4-class \ac{mi}~\cite{Tangermann2012ReviewIV.}.
The main difference to Shallow ConvNet is that EEGNet uses fewer feature maps, spatial separable convolutions, and more pooling layers, which reduces the number of weights as well as the feature map sizes. 
%
%
Its flexibility and small size, however, comes at the cost of significantly lower accuracy, e.g., 67\% for 4-class \ac{mi}. 
Efforts have been made to modify EEGNet by changing the pooling layers and expanding the network achieving 72\% accuracy with subject-specific models~\cite{Uran2019ApplyingAnalysis_short}.

Most of these models are evaluated remotely offline, without considering the possibility to bring the computation closer to the sensors, where the data is acquired. Few studies have shown embedded implementations using traditional \ac{mi}-\acp{bci} with separate feature extractors~\cite{McCrimmon2017PerformancePlatform, Condori2016EmbeddedHand, Belwafi2018AnSystems}, but, to the best of our knowledge, no previous work has demonstrated accurate embedded \ac{mi}-\ac{bci} on low-power \acp{mcu} using \acp{cnn}, which offers better accuracy at lower latency.
In this paper, we propose a \ac{cnn} model based on EEGNet to perform \ac{mi} classification on Physionet dataset~\cite{goldberger2000physiobank}. 
Our proposed model improves the 4-class \ac{mi} accuracy by 6.49\% in global validation while requiring two orders of magnitude less parameters and reducing the memory footprint during inference by a factor of 4.6$\times$ compared to the current \ac{soa} on this dataset. 
In order to target resource-constrained low-power embedded systems, we further study model reduction methods to test the limitations of the proposed model architecture in terms of the sampling frequency, the number of \ac{eeg} channels, and the length of the input signals. We implement two reduced models that are within the resource constraints of two popular low-power \acp{mcu} with ARM Cortex-M4F and M7 and measured the runtime and energy consumption. 
%

\section{Dataset Description}\label{sec:dataset}
\begin{figure}
    \fontsize{8}{10}\selectfont
  \centering
  \includesvg[width=0.95\linewidth]{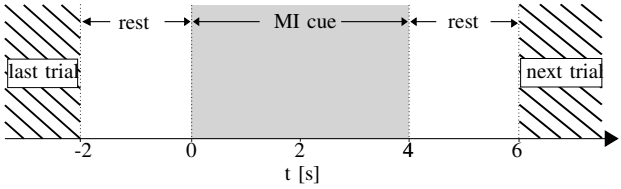}
  \caption{Trial paradigm~\cite{Dose2018AnBCIs} of Physionet EEG Motor Movement/Imagery Dataset.}
  \label{fig:physionet_trial}
    \vspace{-0.3cm}
\end{figure}

We use the publicly available Physionet EEG Motor Movement/Imagery Dataset~\cite{goldberger2000physiobank} containing \ac{eeg} recordings of 109 subjects. 
Four subjects are discarded due to variability in the number of trials, resulting in 105 subjects to be finally used. 
The \ac{eeg} signals were recorded with the BCI2000 system~\cite{Schalk2004BCI2000:System} using 64 channels sampled at 160\,Hz.
The subjects performed motor movement and \ac{mi} tasks, however, in this study we solely focus on the classification of \ac{mi}. 
Every subject participated in three runs for \ac{mi} of left fist (L) against right fist (R), and three runs for \ac{mi} of both fists (B) against both feet (F). 
One run lasts 120\,s and consists of 14 \ac{mi} trials according to the timing scheme shown in Fig.~\ref{fig:physionet_trial}. 
This results in 21 trials per class per subject.
A baseline run provides resting data (0), where the subjects did not receive any cues for 60\,s while having their eyes open.
In order to get trials with resting data, we extract windows of 3\,s from the baseline run. 
As done in~\cite{Dose2018AnBCIs}, we distinguish between 2-, 3-, and 4-class \ac{mi} using L/R, L/R/0, and L/R/0/F MI tasks, respectively. 
\begin{figure}
    \fontsize{8}{10}\selectfont
  \centering
  \includesvg[width=0.9\linewidth]{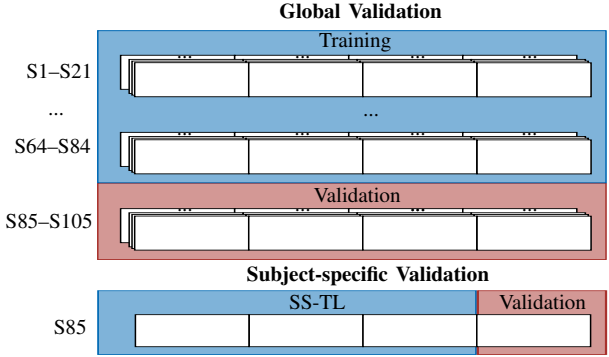}
  \caption{Training and validation on Physionet EEG Motor Movement/Imagery Dataset. Global validation is done via 5-fold \ac{cv} over the subjects, whereas subject-specific validation additionally includes transfer learning (\ac{sstl}) in 4-fold validation on the corresponding subject, e.g., on S85.}
  \label{fig:physionet_testing}
    \vspace{-0.3cm}
\end{figure}

\subsection{Validation methodology} \label{subsec:methodology}
Fig.~\ref{fig:physionet_testing} illustrates the validation methodology inspired by~\cite{Dose2018AnBCIs}, which distinguishes global from subject-specific validation. 
The global validation accuracy is determined by 5-fold \ac{cv} across the subjects, i.e., training on 4/5 of the subjects and validating on the remaining, unseen 1/5 of the subjects.
In \ac{sstl}, the global model is further adjusted by doing transfer-learning on part of the subject's data and validated on the remainder. 
This validation is done with 4-fold \ac{cv} on every subject.

In the example of Fig.~\ref{fig:physionet_testing}, a global model is first trained on subjects S1--S84 and validated on S85--S105, yielding the first fold accuracy of the global validation. 
\ac{sstl} is then applied for S85--S105 on every subject separately in 4-fold \ac{cv}, always starting with the global model from S1--S84.

\section{Methods}\label{sec:methods}
This section introduces the novel embedded \ac{mi}-\ac{bci} model proposed in this paper that matches the memory and complexity constraints of low-power \acp{mcu} with high accuracy. 
We first describe how the compact EEGNet~\cite{Lawhern2018EEGNet:Interfaces} is applied and evaluated on the Physionet Dataset~\cite{goldberger2000physiobank}, and propose methods to further reduce the memory requirements of EEGNet.

\begin{figure*}[ht!]
    \fontsize{8}{10}\selectfont
    \centering
    \includesvg[width=\textwidth]{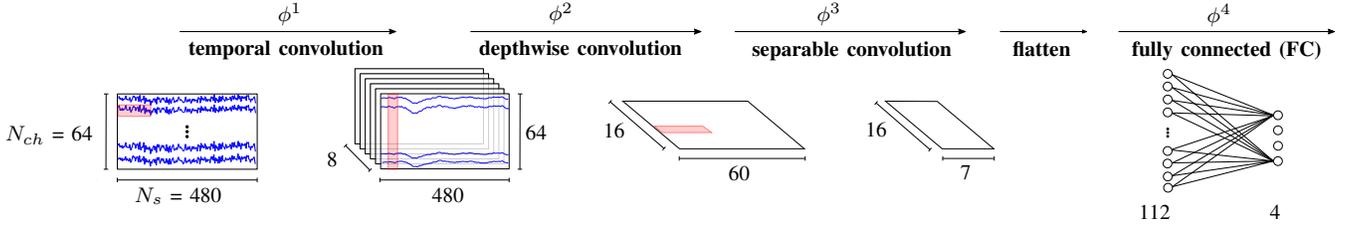}
    \caption{EEGNet~\cite{Lawhern2018EEGNet:Interfaces} in standard configuration for 4-class MI on the Physionet Motor Movement/Imagery dataset. A window of 3\,s ($N_s = 480$ samples) with $N_{ch} = 64$ channels is classified at the time.  
    }
    \label{fig:cnn}
\end{figure*}
\begin{table*}
\caption{Detailed description of EEGNet in MI classification. 
$N_s$ is the number of input samples in time domain, $N_{ch}$ the number of EEG channels, $N_{cl}$ the number of classes, $N_f$ the filter size of first temporal filter, and $N_p$ the pooling length.
For each map, $n$ is the number of filters, $p$ the padding strategy, $k$ the kernel size, and $s$ the stride. 
The last two columns show number of parameters and feature map size of standard configuration, i.e. $N_s=480$, $N_{ch}=64$, $N_{cl}=4$, $N_f=128$, $N_p=8$.
}
    \vspace{-0.15cm}
\label{tab:eegnet}
\begin{tabular}{llccccllll}
                                                &                                  & \multicolumn{1}{l}{}    & \multicolumn{1}{l}{}       & \multicolumn{1}{l}{}                  & \multicolumn{1}{l}{}              &                                                                                                                                      &                                                                                                                           & \multicolumn{2}{c}{Standard Configuration}                                                                              \\ \hline
\multicolumn{1}{|l|}{Layer}                     & \multicolumn{1}{l|}{Type}        & \multicolumn{1}{c|}{n}  & \multicolumn{1}{c|}{p}     & \multicolumn{1}{c|}{k}                & \multicolumn{1}{c|}{s}            & \multicolumn{1}{l|}{Parameters}                                                                                                      & \multicolumn{1}{l|}{Output Shape}                                                                                          & \multicolumn{1}{l|}{Parameters}       & \multicolumn{1}{l|}{\begin{tabular}[c]{@{}l@{}}Feature\\ Map Size\end{tabular}} \\ \hline
\multicolumn{1}{|l|}{\multirow{2}{*}{$\phi^1$}} & \multicolumn{1}{l|}{Conv2d}      & \multicolumn{1}{c|}{8}  & \multicolumn{1}{c|}{same}  & \multicolumn{1}{c|}{$N_f\times 1$}    & \multicolumn{1}{c|}{$1\times1$}   & \multicolumn{1}{l|}{$8N_f$}                                                                                                          & \multicolumn{1}{l|}{\multirow{2}{*}{$N_s \times N_{ch} \times 8$}}                                                        & \multicolumn{1}{l|}{1,024}            & \multicolumn{1}{l|}{\multirow{2}{*}{\textbf{245,760}}}                                   \\ \cline{2-7} \cline{9-9}
\multicolumn{1}{|l|}{}                          & \multicolumn{1}{l|}{BatchNorm2d} & \multicolumn{4}{c|}{-}                                                                                                           & \multicolumn{1}{l|}{32}                                                                                                              & \multicolumn{1}{l|}{}                                                                                                     & \multicolumn{1}{l|}{32}               & \multicolumn{1}{l|}{}                                                           \\ \hline
\multicolumn{1}{|l|}{\multirow{4}{*}{$\phi^2$}} & \multicolumn{1}{l|}{DepthConv2d} & \multicolumn{1}{c|}{16} & \multicolumn{1}{c|}{valid} & \multicolumn{1}{c|}{$1\times N_{ch}$} & \multicolumn{1}{c|}{$1\times0$}   & \multicolumn{1}{l|}{$N_{ch}\cdot16$}                                                                                                 & \multicolumn{1}{l|}{\multirow{4}{*}{$1 \times N_s//N_p \times 16$}}                                                        & \multicolumn{1}{l|}{1,024}            & \multicolumn{1}{l|}{\multirow{4}{*}{960}}                                       \\ \cline{2-7} \cline{9-9}
\multicolumn{1}{|l|}{}                          & \multicolumn{1}{l|}{BatchNorm2d} & \multicolumn{4}{c|}{-}                                                                                                           & \multicolumn{1}{l|}{64}                                                                                                              & \multicolumn{1}{l|}{}                                                                                                     & \multicolumn{1}{l|}{64}               & \multicolumn{1}{l|}{}                                                           \\ \cline{2-7} \cline{9-9}
\multicolumn{1}{|l|}{}                          & \multicolumn{1}{l|}{EluAct}      & \multicolumn{4}{c|}{-}                                                                                                           & \multicolumn{1}{l|}{-}                                                                                                               & \multicolumn{1}{l|}{}                                                                                                     & \multicolumn{1}{l|}{-}                & \multicolumn{1}{l|}{}                                                           \\ \cline{2-7} \cline{9-9}
\multicolumn{1}{|l|}{}                          & \multicolumn{1}{l|}{AvgPool2d}   & \multicolumn{1}{c|}{-}  & \multicolumn{1}{c|}{valid} & \multicolumn{1}{c|}{$N_p \times 1$}   & \multicolumn{1}{c|}{$N_p\times1$} & \multicolumn{1}{l|}{-}                                                                                                               & \multicolumn{1}{l|}{}                                                                                                     & \multicolumn{1}{l|}{-}                & \multicolumn{1}{l|}{}                                                           \\ \hline
\multicolumn{1}{|l|}{\multirow{4}{*}{$\phi^3$}} & \multicolumn{1}{l|}{SepConv2d}   & \multicolumn{1}{c|}{16} & \multicolumn{1}{c|}{same}  & \multicolumn{1}{c|}{$16\times1$}      & \multicolumn{1}{c|}{$1\times1$}   & \multicolumn{1}{l|}{512}                                                                                                             & \multicolumn{1}{l|}{\multirow{4}{*}{$1 \times N_s//N_p//8  \times 16$}}                                                     & \multicolumn{1}{l|}{512}              & \multicolumn{1}{l|}{\multirow{4}{*}{120}}                                       \\ \cline{2-7} \cline{9-9}
\multicolumn{1}{|l|}{}                          & \multicolumn{1}{l|}{BatchNorm2d} & \multicolumn{4}{c|}{-}                                                                                                           & \multicolumn{1}{l|}{64}                                                                                                              & \multicolumn{1}{l|}{}                                                                                                     & \multicolumn{1}{l|}{64}               & \multicolumn{1}{l|}{}                                                           \\ \cline{2-7} \cline{9-9}
\multicolumn{1}{|l|}{}                          & \multicolumn{1}{l|}{EluAct}      & \multicolumn{4}{c|}{-}                                                                                                           & \multicolumn{1}{l|}{-}                                                                                                               & \multicolumn{1}{l|}{}                                                                                                     & \multicolumn{1}{l|}{-}                & \multicolumn{1}{l|}{}                                                           \\ \cline{2-7} \cline{9-9}
\multicolumn{1}{|l|}{}                          & \multicolumn{1}{l|}{AvgPool2d}   & \multicolumn{1}{c|}{-}  & \multicolumn{1}{c|}{valid} & \multicolumn{1}{c|}{$8\times1$}       & \multicolumn{1}{c|}{$1\times8$}   & \multicolumn{1}{l|}{-}                                                                                                               & \multicolumn{1}{l|}{}                                                                                                     & \multicolumn{1}{l|}{-}                & \multicolumn{1}{l|}{}                                                           \\ \hline
\multicolumn{1}{|l|}{\multirow{2}{*}{$\phi^4$}} & \multicolumn{1}{l|}{FC}          & \multicolumn{1}{c|}{4}  & \multicolumn{3}{c|}{-}                                                                                 & \multicolumn{1}{l|}{$(N_s//N_p//8 \cdot 16 +1)N_{cl}$}                                                                                 & \multicolumn{1}{l|}{\multirow{2}{*}{$N_{cl}$}}                                                                            & \multicolumn{1}{l|}{484}              & \multicolumn{1}{l|}{\multirow{2}{*}{4}}                                         \\ \cline{2-7} \cline{9-9}
\multicolumn{1}{|l|}{}                          & \multicolumn{1}{l|}{SoftMaxAct}  & \multicolumn{4}{c|}{-}                                                                                                           & \multicolumn{1}{l|}{-}                                                                                                               & \multicolumn{1}{l|}{}                                                                                                     & \multicolumn{1}{l|}{-}                & \multicolumn{1}{l|}{}                                                           \\ \hline
                                                & \multicolumn{1}{l|}{}            & \multicolumn{4}{l|}{\textbf{Total (inkl. Input Feature Map)}}                                                                    & \multicolumn{1}{l|}{\textbf{\begin{tabular}[c]{@{}l@{}}$\mathbf{672+16N_{ch}+8N_f}$\\ $\mathbf{+(2N_s/N_p +1)N_{cl}}$\end{tabular}}} & \multicolumn{1}{l|}{\textbf{\begin{tabular}[c]{@{}l@{}}$\mathbf{N_s(9N_{ch}+18/N_p)}$\\ $\mathbf{+N_{cl}}$\end{tabular}}} & \multicolumn{1}{l|}{$\mathbf{3,204}$} & \multicolumn{1}{l|}{$\mathbf{277,564}$}                                         \\ \cline{3-10} 
\end{tabular}
\vspace{-0.3cm}
\end{table*}

\subsection{EEGNet on Physionet Dataset}
Fig.~\ref{fig:cnn} shows the architecture of EEGNet for 4-class \ac{mi} on the Physionet Dataset. 
The first 3\,s of the MI cue are used for classification, which is the interval [0\,s, 3\,s] according to Fig.~\ref{fig:physionet_trial}. The input feature map represents the \ac{eeg} signal in time domain with $N_s$=480 samples (3\,s$\times$160\,Hz) and $N_{ch}$=64 \ac{eeg} channels. 
The samples are filtered in the time domain using \mbox{1-D} convolutions, in the spatial domain with depthwise convolutions, in the time-spatial domain with separable convolutions, and finally classified with a fully connected layer.
Furthermore, EEGNet uses \ac{elu} activation and average pooling in the time domain. 

Table \ref{tab:eegnet} gives further insights into the architecture of EEGNet. 
Here, the number of input samples, \ac{eeg} channels, kernel size of the first average pooling are kept variable; they have a direct influence on the number of parameters as well as feature map sizes. 
The last two columns show that EEGNet is indeed very compact: it requires to learn only 3,204 weights, which is two orders of magnitude less than the SoA CNN~\cite{Dose2018AnBCIs}. 
However, large feature maps need to be stored during operation.
Assuming we need to be able to store at least two consecutive feature maps at any time, the maximum number of stored features is the sum of the input and first layer, i.e., $N_s\times N_{ch}+N_s\times N_{ch}\times8=276,480$ features. 

Training and evaluation of the EEGNet are implemented using Keras with TensorFlow (version 1.11) backend. 
The model is trained with Adam optimizer for 100 epochs with batch size of 16 and a fixed learning rate scheduler, setting the learning rate to 0.01, 0.001, and 0.0001 at epochs 0, 20, and 50, respectively. 
In \ac{sstl}, the model is trained for 5 more epochs. 
All training hyperparameters were determined via 5-fold \ac{cv} on the training set of the first fold of the global validation set (i.e., S1--S84) for 4-class \ac{mi}, and kept the same for 2- and 3-class \ac{mi} as well as all reduced EEGNet configurations.

\subsection{Embedded implementation}
\begin{figure}
    \centering
    \includegraphics[width=0.9\linewidth, trim={0cm 3.9cm 0cm 6cm}, clip=true]{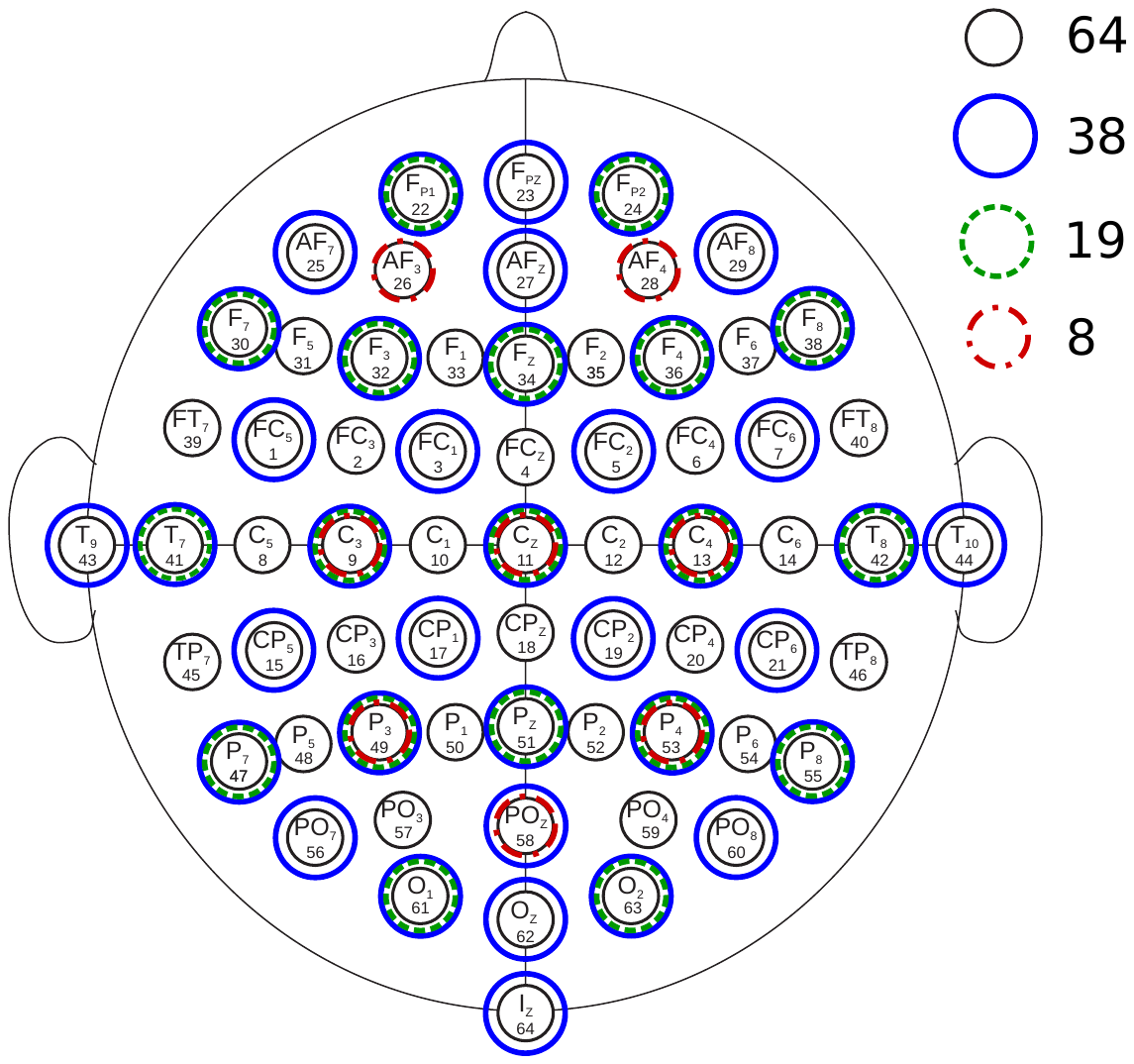}
    \caption{Electrode configurations.}
    \label{fig:channels}
\vspace{-0.4cm}
\end{figure}
For the evaluation on embedded processors, we choose two \acp{mcu} from STMicroelectronics: B-L475-IOT01A2 with an ARM Cortex-M4F processor at 80\,MHz with 128\,KB of SRAM and 1\,MB of Flash memory and STM32F756 Nucleo-144 with an ARM Cortex-M7 processor at 216\,MHz with 320\,KB SRAM and 1\,MB of Flash. Both \acp{mcu} utilize digital signal processors and floating-point units. We then use the X-CUBE-AI expansion package of STM32CubeMX~\cite{cubeai} to deploy the trained models on the selected \acp{mcu}.

Based on Table~\ref{tab:eegnet} and considering 32-bit floating-point numbers, the estimated Flash memory needed for storing the parameters of the model is around 13\,KB, whereas the RAM requirement for the to largest consecutive feature maps is roughly 1.05\,MB. With these configurations, the model can not be executed on the selected low-power \acp{mcu}. As mentioned in the previous subsection, the output of the first layer requires most of the memory. To overcome this limitation, we reduce the input data size by:
\begin{enumerate}
    \item downsampling the \ac{eeg} data in the time domain. 
    \ac{mi} activities cause brain oscillations mainly within the $\mu$ (8--14\,Hz) and $\beta$ (14--30\,Hz) bands~\cite{Schirrmeister2017DeepVisualization}. 
    Some \acp{cnn} have shown to learn temporal filters that cover the gamma (71--91\,Hz) band, however, the main information was still extracted from $\mu$ and $\beta$ oscillations~\cite{Dose2018AnBCIs}. 
    Therefore, we downsample the signals by a factor of $ds$=2 or $ds$=3, which restricts us to maximal oscillations of 40\,Hz or 26\,Hz, respectively. 
    %
    The temporal filter and the pooling kernel sizes are scaled to $N_f=\ceil*{128/ds}$ and $N_p=\ceil*{8/ds}$.
    %
    This way, the network is expected to learn similarly to the original model after the depthwise convolution, independent on the downsampling factor.
    \item using a subset of electrode channels. 
    The BCI2000 system conforms to the 10-10 international system electrode placement with $N_{ch}$=64 electrodes. 
    We reduce the number of electrodes to $N_{ch}$=19 by taking the widely used 10-20 international system electrode placement, from which we exclude A1 and A2. 
    As an intermediate configuration, we add the channels to cover the whole region of the brain equally reaching $N_{ch}$=38 electrodes. 
    We also investigate the case with only 8 electrodes based on the \ac{eeg} headset by Bitbrain.
    Fig.~\ref{fig:channels} shows the 64-, 38-, 19-, and 8-electrodes configurations.
    \item decreasing the time window of the signal used for each classification. We reduce the input signal from 3\,s to 2\,s or 1\,s after the start of the \ac{mi} cue (ref. Fig.~\ref{fig:physionet_trial}).
    Noteworthy, this approach reduces the delay of the system in addition to the model size reduction.
\end{enumerate}
We study the impact on classification accuracy of each reduction approach, testing different configurations to choose the best combination in terms of accuracy and memory footprint for further deployment on both selected \acp{mcu}. 
               
         
\section{Experimental results}\label{sec:results}
This section assesses the proposed methods on the Physionet Motor Movement/Imagery Dataset. 
We measure the classification accuracy as the ratio between correct classified trials over the total number of trials. 

\subsection{Global vs. Subject-specific \ac{mi} Classification}
Table \ref{tab:results} compares the average classification accuracy of the global and subject-specific model of EEGNet with the baseline \ac{cnn} proposed in~\cite{Dose2018AnBCIs}.
In global validation, EEGNet outperforms the baseline \ac{cnn} by 2.05\%, 5.25\%, and 6.49\% on 2-, 3-, and 4-class \ac{mi}, respectively. 
EEGNet does not improve as significantly as the baseline \ac{cnn} when applying SS-TL: the accuracy increases by 1.89\%, 5.00\% and 5.76\% on EEGNet and by 6.11\%, 9.43\%, and 9.93\% on the baseline CNN. 
Due to already high accuracy of EEGNet in global validation, however, the accuracy in subject-specific validation is still 0.82\% and 2.32\% higher than the baseline \ac{cnn} in 3- and 4-class \ac{mi} and only 2.17\% lower in 2-class MI. 
\begin{table}
\vspace{0.05in}
\caption{Classification accuracy (\%) on Physionet EEG Motor Movement/Imagery Dataset using global model (global) or subject-specific model with transfer learning (SS-TL). The top accuracies for each configuration are marked bold.}
\vspace{-0.2cm}
\centering
\label{tab:results}
\begin{tabular}{lcccc} 
\cmidrule(r){1-5} 
& \multicolumn{2}{c}{Dose et al.~\cite{Dose2018AnBCIs}} & \multicolumn{2}{c}{This work} \\ 
\cmidrule(r){2-3} \cmidrule(r){4-5} 
& global & SS-TL & global & SS-TL\\
\cmidrule(r){2-2} \cmidrule(r){3-3} \cmidrule(r){4-4} \cmidrule(r){5-5}
2 classes & 80.38 & \textbf{86.49} & \textbf{82.43} & 84.32 \\
3 classes & 69.82 & 79.25 & \textbf{75.07} & \textbf{80.07}\\
4 classes & 58.58 & 68.51 & \textbf{65.07} & \textbf{70.83}\\
 \cmidrule(r){1-5}
\end{tabular}
    \vspace{-0.4cm}
\end{table}

\subsection{EEGNet Model Reduction}
\begin{table*}
\vspace{0.05in}
\caption{Classification accuracy (\%) using global validation.
The standard EEGNet (Fig. \ref{fig:cnn}) is reduced either by downsampling in time domain, reducing the number of channels, or narrowing the time window for a single classification. 
}
\vspace{-0.2cm}
\centering
\label{tab:reduction}
\begin{tabular}{lcccccccc} 
\cmidrule(r){1-9} 
& Standard & \multicolumn{2}{c}{Downsampling} & \multicolumn{3}{c}{Channel reduction} & \multicolumn{2}{c}{Time window} \\ 
\cmidrule(r){2-2} \cmidrule(r){3-4} \cmidrule(r){5-7} \cmidrule(r){8-9} 
&  & $ds$=2 & $ds$=3 & $N_{ch}$=38 & $N_{ch}$=19 & $N_{ch}$=8 & $T$=2\,s & $T$=1\,s \\ 
\cmidrule(r){3-3} \cmidrule(r){4-4} \cmidrule(r){5-5} \cmidrule(r){6-6} \cmidrule(r){7-7} \cmidrule(r){8-8} \cmidrule(r){9-9} 
2 classes&	82.43&	82.11&	81.97&	81.86&	81.95&	78.07&	81.11&	79.86\\ 
3 classes&	75.07&	74.78&	73.82&	74.12&	72.41&	68.99&	73.45&	71.47\\ 
4 classes&	65.07&	64.81&	64.77&	64.65&	62.55&	58.55&	64.13&	63.51\\ 
\cmidrule(r){1-9}  
\end{tabular}
    \vspace{-0.4cm}
\end{table*}
Table \ref{tab:reduction} studies the impact on the classification accuracy in 2-, 3-, and 4-class \ac{mi} on global validation when reducing EEGNet by temporal downsampling, channel reduction, and narrowing the classified time window. 
Only one reduction approach is applied at a time; the remaining configurations are kept according to the standard EEGNet.

As expected, downsampling has a negligible effect on the accuracy with a maximum decrease among all \ac{mi} tasks of 0.32\% and 1.25\% at downsampling factor $ds$=2 and $ds$=3, respectively.
Even though part of the $\beta$ band is ignored at $ds$=3 due to the cut-off frequency of the anti-aliasing filter at $\approx$26\,Hz, the accuracy does not drop significantly. 
This result confirms that the significant information in this dataset is contained in $\alpha$ and lower $\beta$ bands for the \ac{mi} task. 
When reducing the number of EEG channels to $N_{ch}$=38, the accuracy decreases only marginally by a maximum of 0.95\%. 
However, further reduction to $N_{ch}$=19 and $N_{ch}$=8 significantly affects the performance with a maximum accuracy decrease of 2.66\% and 6.52\%, respectively. 
%
Similar trends can be seen when narrowing down the time window used to do one classification: the accuracy decreases by a maximum of 1.62\% with a temporal window of $T$=2\,s, and by 3.6\% with $T$=1\,s. 
As already mentioned in the previous section, narrowing the time window brings additional advantages in shorter classification delays, and thus, provides a trade-off between accuracy and delay to be chosen by the user. 

\begin{figure}[t]
\centering
 \includegraphics[width=1\linewidth, trim={0, 0, 0, 1cm}, clip=true]{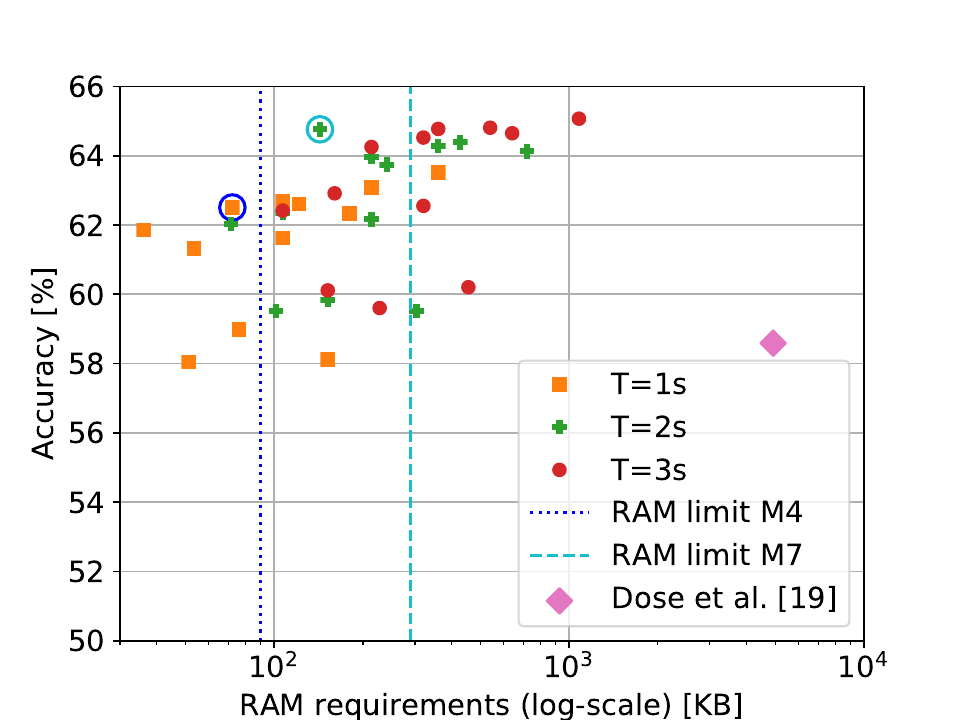}
 \caption{Global accuracy on 4-class \ac{mi} vs. RAM requirements for storing feature maps of reduced configurations of EEGNet. The chosen configurations with highest accuracy while staying below the practical RAM limits are ``Model 1'' (blue circle) with $ds$=3, $N_{ch}$=38, and $T$=1\,s for M4, and ``Model 2'' (cyan circle) with $ds$=3, $N_{ch}$=38, and $T$=2\,s for M7.}%
 \label{fig:acc_vs_memftprnt}
     \vspace{-0.3cm}
\end{figure}

Next, we test all combinations of reduction methods in order to find the best configuration in terms of accuracy vs. memory footprint. 
Fig.~\ref{fig:acc_vs_memftprnt} shows the global 4-class accuracy of all reduction combinations, excluding the $N_{ch}$=8 configuration due to the large drop in accuracy. 
We consider only the memory footprint required to store input and first layer features with 32-bit floating-point representation, since the number of features is two orders of magnitudes higher than the parameters, as pointed out in Table \ref{tab:eegnet}.
As delay might be an additional constraint for model selection, the configurations are marked according to the time window used for classification. 
EEGNet outperforms the baseline \ac{cnn} in most configurations: it has at least 4.6$\times$ lower memory footprint (i.e., $<$1.05\,MB vs. 4.80\,MB), while achieving higher classification accuracy in most cases. 
We select two EEGNet configurations on the pareto-optimal curve, which satisfy the memory constraints of the chosen Cortex-M4F and M7 MCUs. 
Both configurations use a downsampling of $ds$=3 and $N_{ch}$=38 channels; they only differ in the time window choosing $T$=1\,s at 62.51\% accuracy and 72\,KB RAM requirements for Cortex-M4, and $T$=2\,s at 64.76\% with 143\,KB for Cortex-M7.
This corresponds to an accuracy loss of 2.51\% at 15$\times$ model reduction when operating EEGNet on the M4, and 0.31\% loss at 7.6$\times$ model reduction on the M7, compared to the standard EEGNet configuration. 
We name them ``Model 1'' and ``Model 2'', respectively, in shorthand.



\subsection{MCU Implementation}

We deploy the selected models using STM32CubeMX v5.3 with X-CUBE-AI 5.0.0 package extension on STM32L475VG \mbox{B-L475E-IOT01A} with an ARM Cortex-M4F processor and STM32F756ZG Nucleo-144 with an ARM Cortex-M7 processor and measure the power consumption with a Keysight N6705C power analyzer. In order to have optimal performance, we enable the core instruction and data caches and ART accelerator sub-system to speed up instruction fetch accesses. We deploy Model 1 on both Cortex-M4F and Cortex-M7 for comparison, and Model 2 only on Cortex-M7 due to memory constraints. 

The Cortex-M7 offers the highest performance in ARM Cortex-M processor family. In fact, as can be seen in Table \ref{tab:M4 vs M7}, it takes around 6 cycles per \ac{macc} operation, which is around 1.8$\times$ faster than the Cortex-M4F. However, the power consumption is 3.1$\times$ higher than Cortex-M4F at the same frequency of 80\,MHz. Cortex-M7 can run up to 216\,MHz, which reduces the latency of Model 1 by a factor of 4.9$\times$ compared to Cortex-M4F at a price of almost 2$\times$ more energy consumption. Model 2 has the lowest accuracy loss (i.e., 0.31\%) after model reduction, but it can fit only into the Cortex-M7 processor. Running at the highest frequency, it takes around 44\,ms and 18.1\,mJ per inference.

\begin{table}[]
\centering
\caption{ARM Cortex-M4F vs. M7 on 4-class MI. Both Model 1 (62.51\% accuracy) and Model 2 (64.76\%) use a downsampling of $ds$=3 and $N_{ch}$=38 channels, the former has time window $T$=1\,s, while the latter $T$=2\,s.}
\vspace{-0.2cm}
\label{tab:M4 vs M7}

\setlength{\tabcolsep}{4.3pt}
\begin{tabular}{@{}lcccc}
\midrule
& \multicolumn{3}{c}{Model 1} & Model 2 \\
\cmidrule(r){2-4} \cmidrule{5-5}
\ac{macc} & \multicolumn{3}{c}{761,956} & 1,509,220 \\
ROM size {[}KB{]}        & \multicolumn{3}{c}{6.61}     &   7.12        \\
RAM size {[}KB{]}      & \multicolumn{3}{c}{70.27}    &   139.20      \\
\cmidrule(r){2-4} \cmidrule{5-5}
\multirow{2}{*}{}& M4F & \multicolumn{2}{c}{M7} & M7 \\
\cmidrule(r){2-2} \cmidrule(r){3-4} \cmidrule{5-5}
& @80\,MHz & @80\,MHz & @216\,MHz & @216\,MHz \\
\cmidrule(r){2-2} \cmidrule(r){3-3} \cmidrule(r){4-4} \cmidrule{5-5}
Cycles/\ac{macc}     & 10.59     &5.77     & 5.78    &   6.27        \\
Power {[}mW{]}  @\,3.3V  & 42.44 & 131.41      & 412.76          &   413.06       \\
T/inference {[}ms{]} & 100.84 & 54.99   & 20.40         &   43.81     \\
En./inference {[}mJ{]}          & 4.28 & 7.23        & 8.42           &  18.1 \\
\midrule
\end{tabular}
\vspace{-0.4cm}
\end{table}

\section{Conclusion}\label{sec:conclusion}
\vspace{-0.1cm}
We propose an embedded model based on EEGNet for low-power \ac{mi}-\acp{bci}. The proposed model achieves 2.05\%, 5.25\%, and 6.49\% higher accuracy than the \ac{soa} CNN on \mbox{2-,} \mbox{3-,} and 4-class \ac{mi} classification, while requiring two orders of magnitude less memory for storing model parameters and 4.6$\times$ less memory for feature maps during inference execution. 
We reduce the input feature map by downsampling in the temporal and spatial domain as well as narrowing down the time window and relax the memory requirements by 7.6$\times$ at 0.31\% accuracy loss, and by 15$\times$ at 2.51\% loss. 
We demonstrate the performance of the proposed models on two commercial \acp{mcu}. In particular, the implemented models execute in around 44\,ms consuming 18.1\,mJ per inference on an ARM Cortex-M7 and in 101\,ms using 4.28\,mJ on an ARM Cortex-M4F processor, making them suitable for a battery-operated real-time wearable system to continuously perform online MI classification.

\section*{Acknowledgment}
\vspace{-0.1cm}
This project was supported in part by the Swiss Data Science Center PhD Fellowship under grant ID P18-04 and in part by ETH Research Grant 09 18-2.

\bibliographystyle{IEEEtran}
\bibliography{references.bib,bci_recording.bib,bib.bib}

\end{document}